\documentclass[journal]{IEEEtran}
\usepackage[latin1]{inputenc}
\usepackage{url}
\usepackage{subfig}
\usepackage{ragged2e}
\usepackage{multirow}
\usepackage{siunitx}
\usepackage{algorithm}
\usepackage{algpseudocode}
\usepackage{amsmath} % for 'bmatrix' environment
\ifCLASSINFOpdf
   \usepackage[pdftex]{graphicx}
\else
\fi

\begin{document}

\title{Testbed for Connected Artificial Intelligence using Unmanned Aerial Vehicles and Convolutional Pose Machines}

\author{Diego Dantas\IEEEauthorrefmark{1},~Carnot Braun\IEEEauthorrefmark{1},~Kaio Forte\IEEEauthorrefmark{1}, Fl\'avio Brito\IEEEauthorrefmark{1},~Andrey Silva\IEEEauthorrefmark{1},~Silvia Lins\IEEEauthorrefmark{3}, Neiva Linder\IEEEauthorrefmark{2} and~Aldebaro~Klautau\IEEEauthorrefmark{1}

\IEEEauthorblockA{\IEEEauthorrefmark{1}Federal University of Par\'a, Bel\'em, PA, Brazil\\
                                       \{diego.figueiredo, carnot.filho, kaio.forte, flavio.brito\}@itec.ufpa.br, \{andreysilva, aldebaro\}@ufpa.br}

\IEEEauthorblockA{\IEEEauthorrefmark{2}Ericsson Research, Kista, Sweden\\
                                       \{neiva.linder\}@ericsson.com}

\IEEEauthorblockA{\IEEEauthorrefmark{3}Ericsson Research Brazil, S\~ao Paulo, Brazil\\
                                        \{silvia.lins\}@ericsson.com}

}

\maketitle

%%%%%%%%%%%%%%%%%%%%%%%%%%%%%%%%%%%%%%%%%%%%%%%%%%%%%%%%%%%%%%%%%%%%%%%%%%%%%%%%
\begin{abstract}

Unmanned Aerial Vehicles (UAVs) became very popular in a vast number of applications in recent years, especially drones with computer vision functions enabled by on-board cameras and embedded systems. Many of them apply object detection using data collected by the integrated camera. However, several applications of real-time object detection rely on Convolutional Neural Networks (CNNs) which are computationally expensive and processing CNNs on a UAV platform is challenging (due to its limited battery life and limited processing power). To understand the effects of these issues, in this paper we evaluate the constraints and benefits of processing the whole data in the UAV versus in an edge computing device. We apply Convolutional Pose Machines (CPMs) known as OpenPose for the task of articulated pose estimation. We used this information to detect human gestures that are used as input to send commands to control the UAV. The experimental results using a real UAV indicate that the edge processing is more efficient and faster (w.r.t battery consumption and the delay in recognizing the human pose and the command given to the drone) than UAV processing and then could be more suitable for CNNs based applications.

\end{abstract}

%%%%%%%%%%%%%%%%%%%%%%%%%%%%%%%%%%%%%%%%%%%%%%%%%%%%%%%%%%%%%%%%%%%%%%%%%%%%%%%%
\section{Introduction}

Unmanned Aerial Vehicles (UAVs) are becoming a widely used platform with a broad range of sophisticated applications such as data offloading for mobile networks~\cite{cheng2018uav}, wireless communication~\cite{wu2018joint}, hyperspectral remote sensing~\cite{aasen2018quantitative,zhong2018mini}, intelligent transportation systems~\cite{maiouak2019dynamic, sharma2018lorawan}, and disaster monitoring~\cite{kanzaki2019uav, man2019application}. To develop fully autonomous systems in such applications, is crucial that the drone has an automatic understanding in real time of the visual data collected by the camera~\cite{lee2017real}. However, a key challenge in the deployment of the object detection task is the limited battery (which affects the  flight  time) and the computational power to process the  visual data in a reasonable time to make critical decisions, such  as object avoidance and navigation~\cite{plastiras2018efficient}. Moreover, the task of object detection can face challenges such as image noise and low resolution. 

Most of state-of-the-art algorithms for object detection are based on Convolutional Neural Networks (CNNs) such as the works conducted by~\cite{he2017mask, redmon2018yolov3, pedoeem2018yolo, zhang2019slimyolov3}. Running CNNs locally on UAVs requires a high processing hardware and usually the high power consumption of Graphics  Processing  Units (GPUs) in (low power) UAVs is prohibitive~\cite{plastiras2018efficient}. 

The edge/cloud processing and the split of the CNNs through the network~\cite{lee2017real, teerapittayanon2017distributed, tdedrodud2019uav} is receiving special attention as an alternative to the local processing on UAV. In a scenario with edge/cloud processing without split, the data collected by the UAV is sent to an edge node or to a cloud server for analysis and the result is sent back to the UAV. In the split CNNs scenario, the processing is realized over distributed computing hierarchies (consisting of the cloud, edge and distributed end devices). Both cases enable the UAV to save energy by off-loading computational operations. However, depending on the network conditions and the processing power of the edge/cloud nodes, the wireless transmission of the video can add significant delays to the system. Therefore, for delay-sensitive applications, this kind of solution may not be suitable. Thus, local processing could be recommended~\cite{kyrkou2018dronet}.

Many  UAV  studies  are being conducted related to the deployment of UAVs with high computational cost tasks due to the increasing necessity of solving complex problems using machines. In~\cite{motlagh2017uav}, the UAV is used for crowd surveillance based on face recognition. In~\cite{messous2019game}, the drones have to carry-out pattern recognition and video preprocessing, both considering the possibility of local processing or offloading the intensive computation tasks to a Mobile Edge Computing (MEC) node. In~\cite{lee2017real} the authors propose moving the computation to an off-board computing cloud, while keeping low-level object detection and short-term navigation onboard, comparing the Faster R-CNNs~\cite{ren2015faster} algorithm with YOLOv3~\cite{redmon2016you} and SSD~\cite{liu2016ssd}. In the work presented in~\cite{tdedrodud2019uav} the authors proposed different scenarios to process the data collected by the drone and a simulation is used to compare two neural networks accuracy: Mask R-CNN~\cite{he2017mask} and YOLOv3, running exclusively in the drone.

In this work, in contrast, we use OpenPose~\cite{cao2018openpose} Neural Network, to detect and classify people's actions, using a drone to capture the video input. The drone uses the OpenPose output to identify if a person raised his right hand and it takes off if this specific movement is detected. Two scenarios were evaluated for the processing of OpenPose. In the first scenario, the drone processes using the Intel Computer Neural Stick 2~\cite{intelNS}. In the second scenario, video data is sent through the network to a computer which processes the data and returns to the drone the obtained results. In both scenarios, we evaluate the action recognition time and the power consumption of the drone. The results shows that the edge processing is more suitable to this end. However, we highlight that the network conditions are excellent, i.e., no drops or congestion occurs during the experiment. Besides, the edge node is only processing a single UAV application. Therefore, the results can change significantly if more than one UAV is sending data to the same edge or if a background traffic is present.

The main contributions of this paper are listed below.
 
\begin{itemize}
 \item  We proposed a new algorithm that uses a human pose estimator machine output as its input to detect the human gestures (raised arm) to control the drone (takeoff).
 \item We successfully developed a proof of concept using a real hardware platform integrated with the power measurement module and an USB VPU that can serve as baseline for future investigations, such as the CNNs network split. 
  \item We evaluate, through extensive field tests, the action recognition time and the battery consumption of the drone running
  in the two case studies (whole processing locally at the UAV and at the edge) comparing the most appropriated one for this application. Moreover, we evaluate the processing time of each stage when edge processing is being used (network delay, encoding delay, frame extraction and raw processing). 
\end{itemize}
 
 The rest of this work is organized as follows: Section~\ref{sec:experiments} describes the experiment configuration. Section~\ref{sec:results} provides the evaluation of the obtained results, and Section~\ref{sec:conclusions} concludes the work.

\section{Testbed Description}
\label{sec:experiments}

The testbed consists of using the OpenPose to extract the human pose from 2D images and detect certain human movements to control a drone. The action recognition time is calculated based on the total time needed for one movement to be detected. Everything runs on a real drone but the motors were turned off so the drone doesn't fly and only receives commands as if it were flying. In the following, we describe in details all the experiment parameters.

\subsection{Use cases}
\label{sec:use_cases}
\begin{figure}[htbp]
    \centering
    \includegraphics[width=0.45\textwidth]{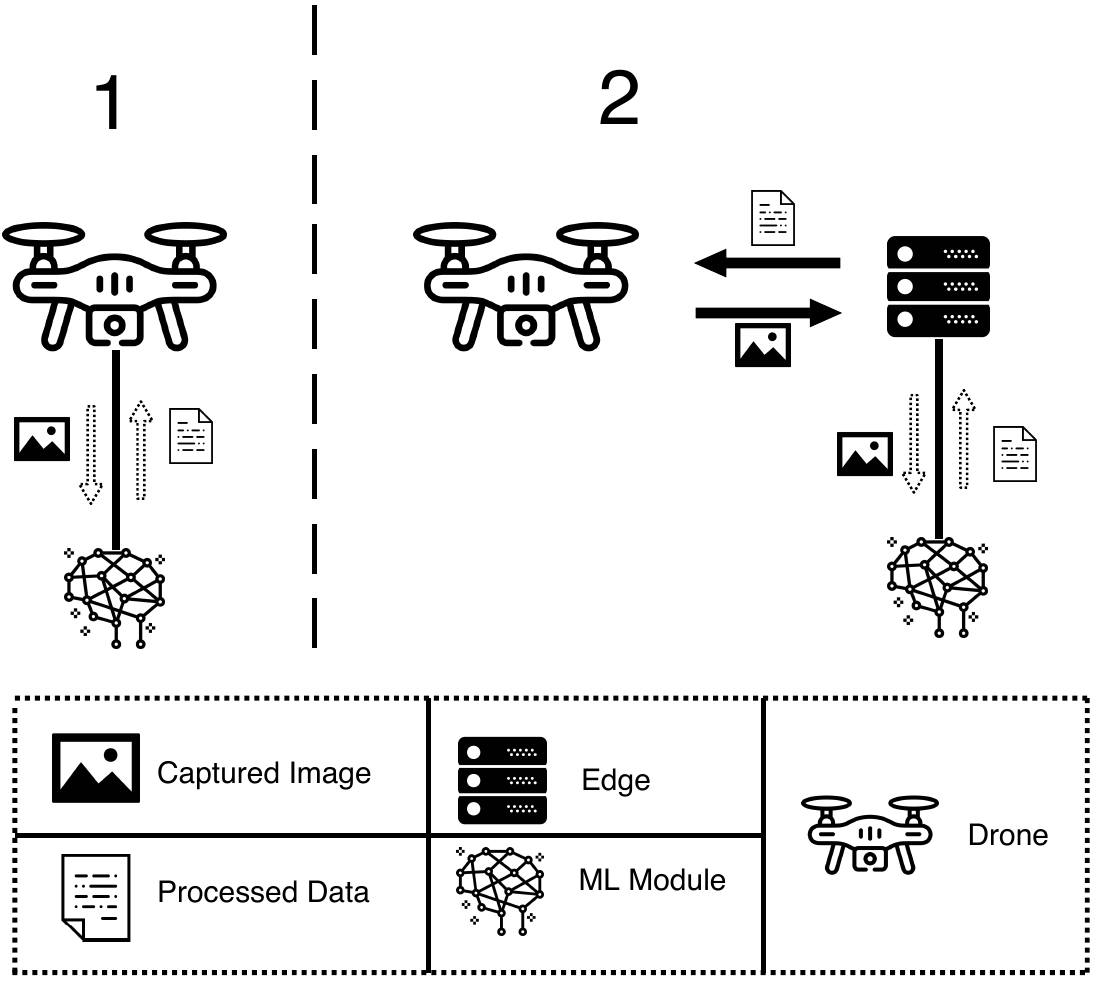}
    \caption{User cases scenarios.}
    \label{fig:usecases}
\end{figure}

We developed the testbed to use the drone for comparing action recognition time and battery consumption for the use cases presented in Fig.~\ref{fig:usecases}. 

Both scenarios are described as follow:
\begin{itemize}
    \item \textbf{CNNs running exclusively in the drone (Fig.~\ref{fig:usecases}, scenario 1):} Applications using CNNs are computationally demanding which makes their use in low cost UAVs infeasible. However, nowadays there are already small VPUs that can be easily carried by drones, such as the Intel Neural Compute Stick 2~\cite{intelNS}, which is a very competitive solution available on the market, where the whole data processing can happen in real-time on companion computers with low power processing capabilities using a USB port. 
    
    \item \textbf{CNNs running in the edge (Fig.~\ref{fig:usecases}, scenario 2):} In this scenario, the drone send the data directly to the edge. The data processing is performed, and the output is sent back to the drone. Under those circumstances, the computational cost for UAVs is very low. The network conditions will be ideal, which means that network delay time evaluation can not be generalized but still giving reasonable estimation.

\end{itemize}

Table~\ref{tab:scenario_configurations} describes the configuration of each scenario developed.

\begin{table}[!h]
\centering
\begin{tabular}{l|l|l|}
\cline{2-3}
                          & Local                                                                & Edge                                                                                                           \\ \hline
\multicolumn{1}{|l|}{CPU} & \begin{tabular}[c]{@{}l@{}}Intel Atom x7-Z8750 \\ @ 2.56 GHz x 4\end{tabular} & \begin{tabular}[c]{@{}l@{}}Intel Core i7-7700 \\ @ 3.60 GHz x 8\end{tabular} \\ \hline
\multicolumn{1}{|l|}{GPU} & None                                                                      & Nvidia GeForce RTX 2070                                                                        \\ \hline
\multicolumn{1}{|l|}{VPU} & \begin{tabular}[c]{@{}l@{}}Intel Movidius \\ Myriad X\end{tabular}        & None                                                                                                                 \\ \hline
\multicolumn{1}{|l|}{RAM} & 4 GB                                                                      & 32 GB                                                                                                                \\ \hline
\end{tabular}
\caption{Configuration for the two scenarios.}
\label{tab:scenario_configurations}
\end{table}

\subsection{Drone details}

We used the drone known as Intel Aero Drone~\cite{intelRTFdrone}, which is flexible and is able to connect a wide variety of sensors and peripherals. %In addition to several fixed-function interfaces: USB3.0 OTG, micro-HDMI, CSI-2 (MIPI), M.2 for SSD, M.2 for LTE, micro-SD, and HSUART, the Compute Board is designed with a connector expansion IO that exposes 6 processor GPIOs, 28FPGA GPIOs, and 5 FGPA analog sense (ADC) inputs and a built-in version of Yocto as its operating system (but capable of changes). 
The drone can communicate with other devices such as smartphones or laptops over a standard WiFi network, which gives to it the ability to transfer commands via another device. The camera resolution is 640x480 pixels, with each frame having ~45Kb of size. 

%For the first scenario, the drone is equipped with one VPU, Intel Neural Compute Stick 2, that will make possible the local processing of the neural network with decent performance. For the second scenario, the drone communicates with a computer that composes the edge via one wireless network, sending video frames and receiving OpenPose output. 
\subsection{Setup for power measuring}

To obtains the battery consumption estimation, we developed a small circuit on a breadboard (since the drone is staying on the ground) using an Arduino Mega 2560~\cite{ArduinoMega}, a voltage divider and the ACS712~\cite{ACS712} current sensor module connected (in parallel and series, respectively) with the battery. The voltage divider consists of 470K\si{\ohm} and a 120K\si{\ohm} resistors with 5\% accuracy for solid measurement. The ACS712 uses the Hall effect to transform the current into a voltage value that can be read by the Arduino.

\begin{figure}[htbp]
    \centering
    \includegraphics[width=0.45\textwidth]{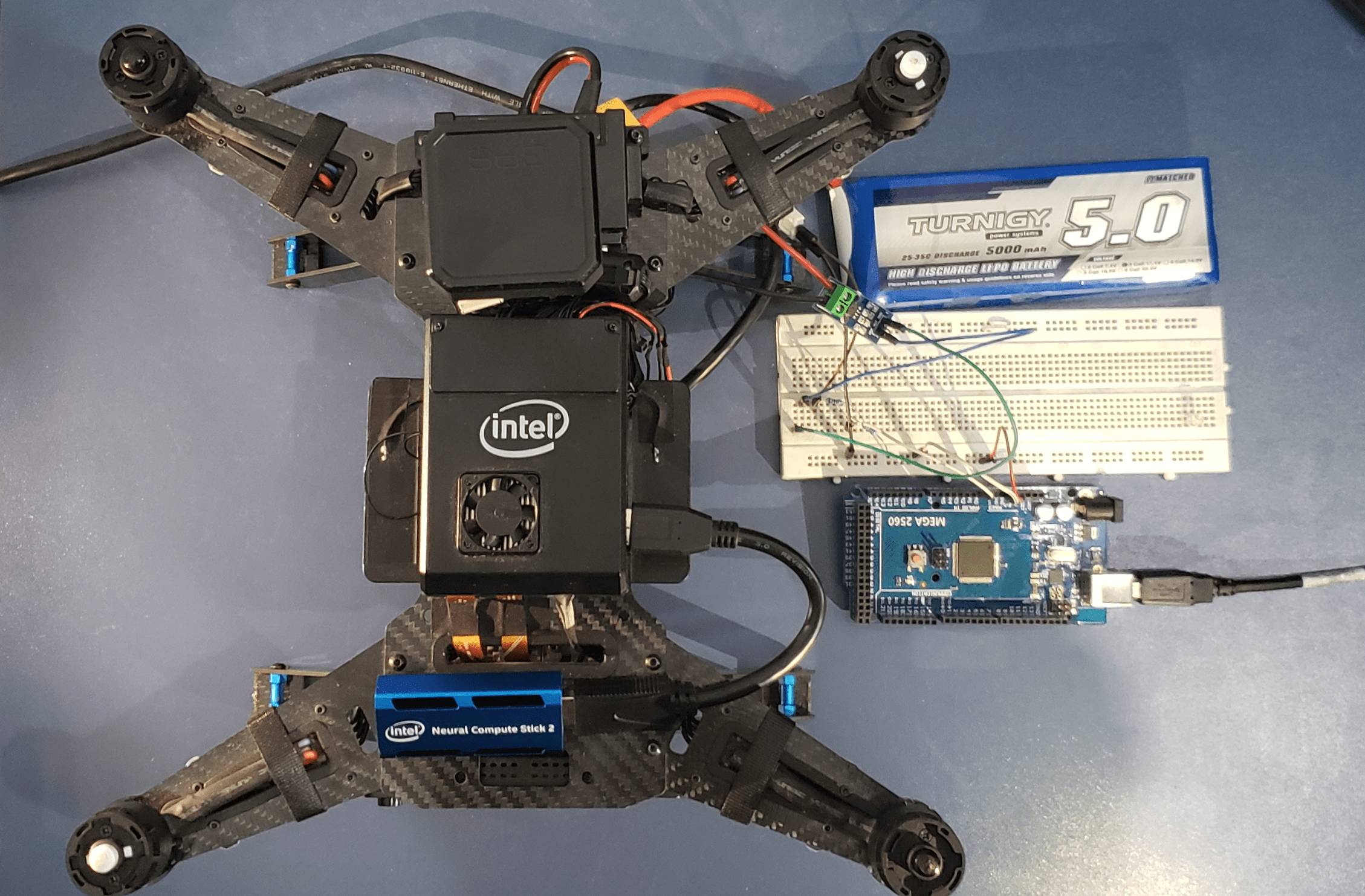}
    \caption{Equipment used in the testbed.}
    \label{fig:equip_used_in_simulation}
 \end{figure}

Both sensors had their outputs connected to the analog input ports of the microcontroller, and then the voltage and current values are measured making the appropriate conversion using the constants related to each sensor. Finally, these measured values were transmitted through the Arduino serial port connected to a computer running Linux, where they were saved for further analysis. Fig.~\ref{fig:equip_used_in_simulation} shows the equipment used for simulations.

\subsection{Movement detector}

\begin{figure}[htbp]
    \centering
    \includegraphics[width=0.45\textwidth]{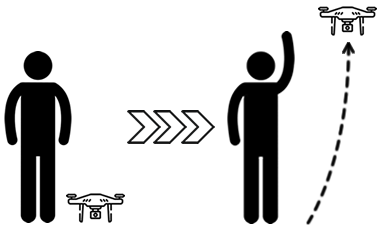}
    \caption{Experiment action controlling the drone.}
    \label{fig:SimulationIlustration}
\end{figure}

The movement selected for the experiment is one person raising its right arm. After detecting the movement, the drone should takeoff. Fig.~\ref{fig:SimulationIlustration} shows an illustration of the expected drone reaction after the human command.

\begin{figure}[htbp]
 \quad \quad  \includegraphics[width=0.45\textwidth]{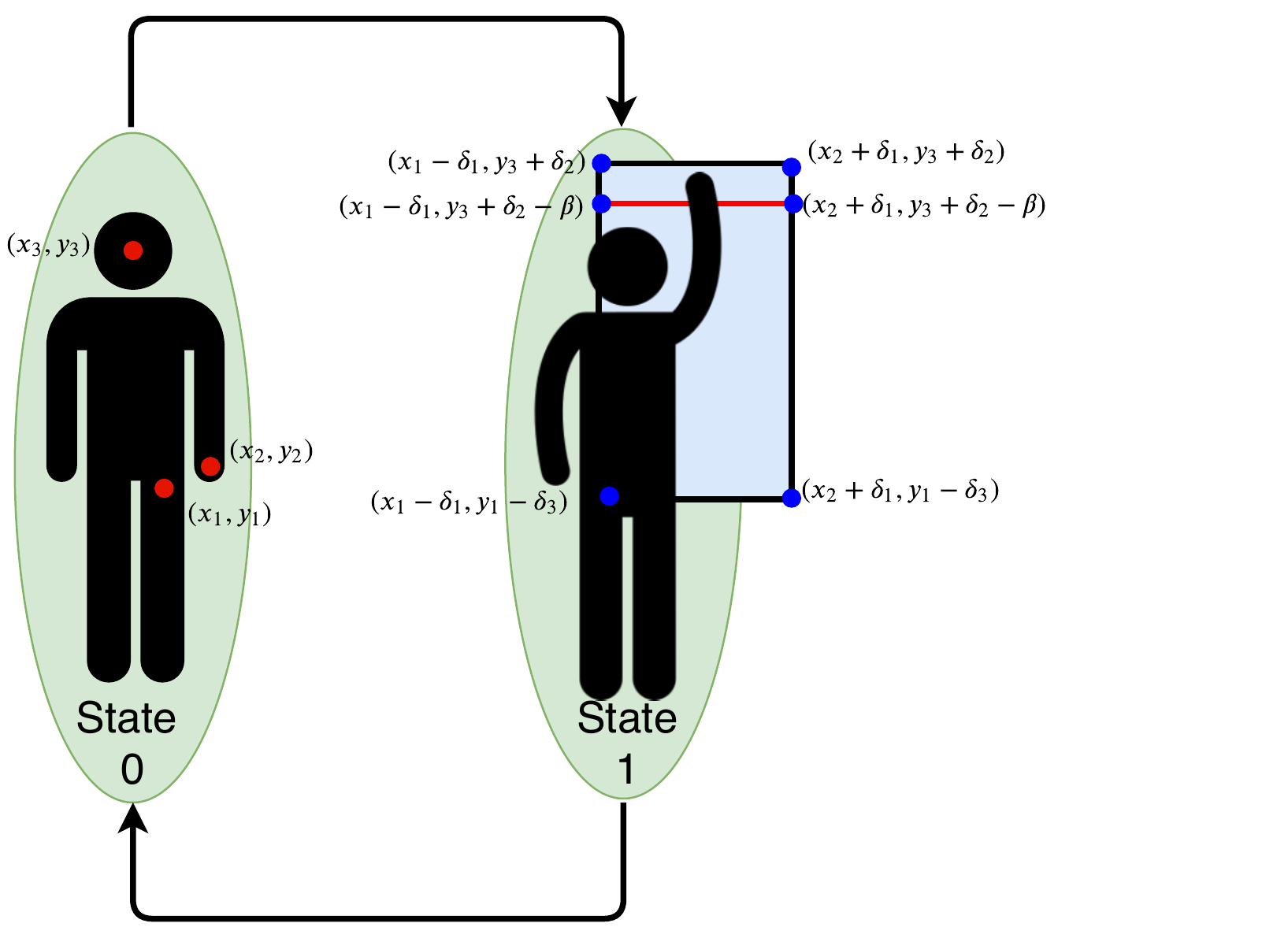}
    \caption{Points of interest for the algorithm.}
    \label{fig:rectangle_creation}
\end{figure}

The algorithm to detect the movement is based in the OpenPose output. First, it predicts the coordinate points where the arm would be when the arm is raising. To this end, the knowledge of the coordinates of the head and the right leg is necessary. Every point will follow a trajectory up and be within a specific region that is possible to be characterized as shown in Fig.~\ref{fig:rectangle_creation}.

\begin{algorithm}
\caption{Main procedure for movement detector}\label{MovementDetector}
\begin{algorithmic}[1]
\State $state\gets 0$
\Procedure{Process Pose}{$HumanPose$}
    \If{$state = 0$} \Comment{Waiting for initial position}
        \State $state\gets initial\_position(HumanPose)$ %\Comment{Checking points position}
        \If{$state = 1$} \Comment{Initial position verified}
            \State $calculate\_box(HumanPose)$ %\Comment{Calculate threshold box}
        \EndIf
    \Else \Comment{Action is being performed}
        \State $state\gets trajectory\_check(HumanPose)$
        \If{$state = 2$} \Comment{Action was performed}
            \State $state\gets 0$ \Comment{Back to initial state}
            \State \textbf{return} $true$ 
        \EndIf
    \EndIf
    \State \textbf{return} $false$ \Comment{Action was not performed}
\EndProcedure
\end{algorithmic}
\label{alg:0}
\end{algorithm}

The \emph{State} $0$ is the initial state, and it waits for the person to be standing straight with the right arm down. This condition can be verified by checking two different euclidean distances. The first one is between the point located in the right thigh and the point located in the right hand. In Fig.~\ref{fig:rectangle_creation}, the right thigh coordinates are defined as $(x_1, y_1)$ and the right hand as $(x_2,y_2)$. Then, the euclidean distance $\alpha_1$ can be found using Equation~\ref{eq:alpha1}.

\begin{equation}
    \alpha_1 = \sqrt{(x_2 - x_1)^2 + (y_2 - y_1)^2} 
    \label{eq:alpha1}
\end{equation}

The second euclidean distance necessary is between the head point and right hand point. In Fig~\ref{fig:rectangle_creation}, these coordinates are $(x_3, y_3)$ and $(x_2, y_2)$ respectively and the distance $\alpha_2$ can be found using Equation~\ref{eq:alpha2}.

\begin{equation}
    \alpha_2 = \sqrt{(x_3 - x_2)^2 + (y_3 - y_2)^2} 
    \label{eq:alpha2}
\end{equation}

After calculate the distances $\alpha_1$ and $\alpha_2$, the system determines if the right arm is down by checking the conditions $\alpha_1 < \beta_1$ and $\alpha_2 < \beta_2$, where $\beta_1$ and $\beta_2$ are constant values. If both conditions are satisfied, then the system goes from \emph{State} $0$ to \emph{State} $1$.

%In \emph{State} $1$, the first step is to design the blue rectangle showed in Fig.~\ref{fig:rectangle_creation}. The coordinates of the left side of the rectangle is the head and thigh coordinates (points $(x_1,y_1)$ and $(x_3,y_3)$ respectively). The coordinates of the rectangle as well as the region delimited by the red line showed in Fig.~\ref{fig:rectangle_creation} are: $\delta_1$, $\delta_2$ and $\delta_3$ which are defined in Equations~\ref{eq:delta_1},~\ref{eq:delta_2},~\ref{eq:delta_3}. $\beta$ is a constant value. 

In \emph{State} $1$, the first step is to design the blue rectangle and Fig~
\ref{fig:rectangle_creation} shows the coordinated of the rectangle design. Basically, we use 
the three points $(x_1, y_1)$, $(x_2, y_2)$ and $(x_3, y_3)$ founded in \emph{State 0} and more 
three variables named $\delta_1$, $\delta_2$ and $\delta_3$ which is calculated using Equations~
\ref{eq:delta_1},~\ref{eq:delta_2} and~\ref{eq:delta_3}. The fourth variable is $\beta$ which is 
a constant value. Using all these variables, the rectangle can be created as Fig~
\ref{fig:rectangle_creation} illustrates.

\begin{equation}
    \delta_1 = \frac{|x_1 - x_2 |}{2}
    \label{eq:delta_1}
\end{equation}

\begin{equation}
    \delta_2 = \frac{|y_1 - y_3 |}{2}
    \label{eq:delta_2}
\end{equation}

\begin{equation}
    \delta_3 = \frac{|y_1 - y_3 |}{4}
    \label{eq:delta_3}
\end{equation}

After the rectangle creation, the current right-hand position $(x_2, y_2)$ is stored and a loop starts. At each loop iteration, the system receives the current right hand position $(x_{2c}, y_{2c})$. With the new current position $(x_{2c}, y_{2c})$ and the last position $(x_2, y_2)$, three observations are performed for checking the completion (or not) of the movement. 
The first observation is to verify if the current point $(x_{2c}, y_{2c})$ is out of rectangle delimited by the points. The second is to certify if the current point $(x_{2c}, y_{2c})$ is within the rectangle region, and if the current point $y_{2c}$ is higher than $y_2$. Then, the last observation is to verify if the person's hand is within the interest region (rectangle region), if positive, the algorithm successfully detected the movement. The complete operation the proposed action recognition scheme is detailed in Algorithm~\ref{alg:0}, where the functions \textit{calculate\_box} and \textit{trajectory\_check}, are responsible for the rectangle design and the verification of the right-hand trajectory. The algorithm also needs to ensure that the person is standing and not in another position (such as lying down), and it can be done by ensuring that the x coordinates of the head and thigh are closer. 

%if it is out of the region or the current point is bellow relatively to the last point of the trajectory, it means that the desired movement was not being executed and the state is changed to 0, if the current point reaches the upper part of the defined region up the red line from the figure, it means that the movement was done and the state is changed to 0 back again.

%All operations of this algorithm have low computational cost because 
%Regarding to the computational cost, the whole algorithm~\ref{MovementDetector} has a time complexity of $O(1)$ and, therefore, the impact in the experiment performance can be unconsidered.

\begin{figure}[htbp]
   \centering
   \includegraphics[width=0.45\textwidth]{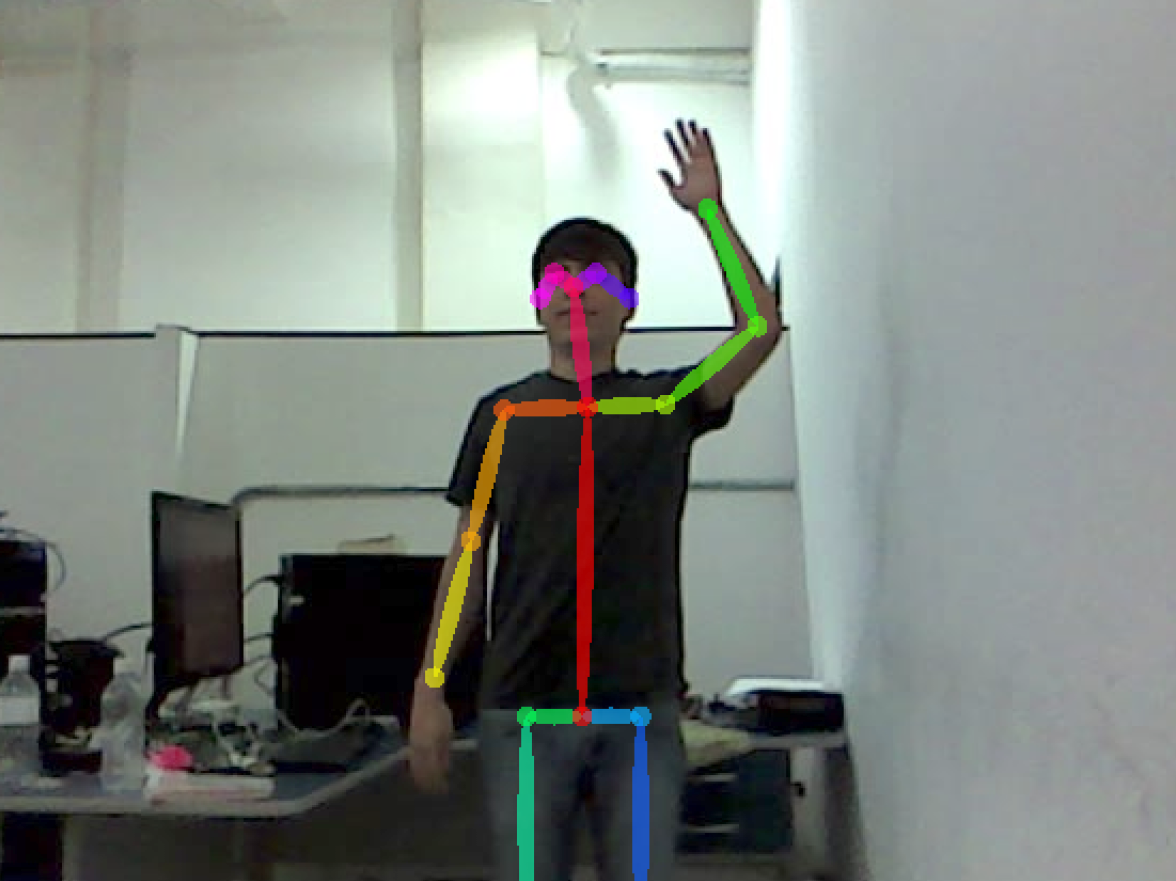}
   \caption{A scene of the video processed by OpenPose used in the experiment.}
   \label{fig:diego_pose}
\end{figure}

\section{Results}
\label{sec:results}

In this section, we evaluate the action recognition time and the battery consumption performance of the drone application in both scenarios proposed for the testbed. All experiments were conducted in an indoor environment. For a fair comparison of the battery usage, the energy consumption was evaluated until the drone received the takeoff signal thus not considering the power consumption from motors. So we can consider that the achieved result represents the use of one battery exclusively for the companion computer.

The camera video is passed as input to the algorithm. Matrix A, B, and the Equation~\ref{matrix_sum} show the acquisition process of the results for both power consumption and recognition time. Each element $a_{i,j}$ of the matrix $A$ represent the time, or power consumption, needed for one action be recognized and therefore we will call it one iteration. Each row in matrix $A$ represents a sample composed of successive 50 iterations, that is, the battery of the drone was not recharged affecting the power consumption analysis. We take the mean of each column of matrix $A$ with Equation \ref{matrix_sum} representing the average value for both parameters being analyzed at a given level of the battery, and matrix B is generated with these values.

\[
A = \begin{bmatrix} 
    a_{1,1} & a_{1,2} & a_{1,3} & \dots & a_{1,50} \\
    a_{2,1} & a_{2,2} & a_{2,3} & \dots & a_{2,50} \\
    a_{3,1} & a_{3,2} & a_{3,3} & \dots & a_{3,50} \\
    a_{4,1} & a_{4,2} & a_{4,3} & \dots & a_{4,50} \\
    a_{5,1} & a_{5,2} & a_{5,3} & \dots & a_{5,50} \\
    \end{bmatrix}
\]

\[
B = \begin{bmatrix}
    \label{resultmatrix}
    f(1) & f(2) & f(3) & \dots & f(50) \\
    \end{bmatrix}
\]

\begin{equation}
	\label{matrix_sum}
    f(j)=\sum_{i=1}^{5}a_{ij}
\end{equation}

\begin{figure}[htbp]
   \centering
  \includegraphics[width=0.45\textwidth]{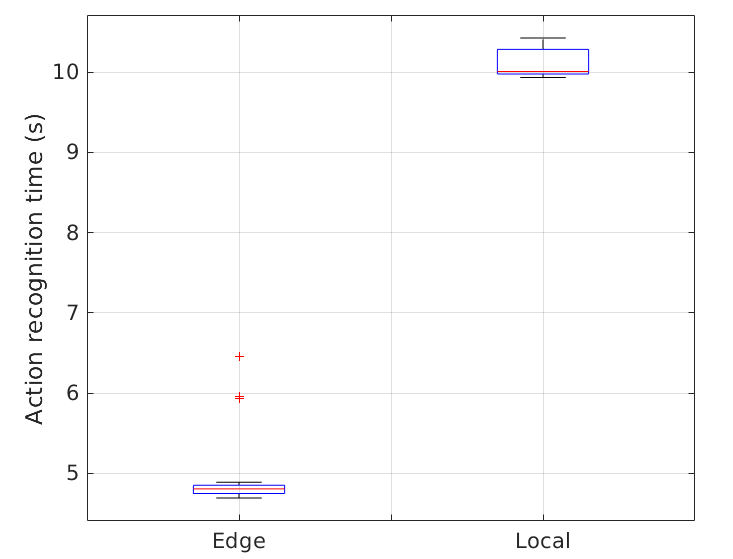}
  \caption{Action recognition time for each approach.}
   \label{fig:latency}
\end{figure}

The action recognition time may seem impractical in both use cases for real time systems but this delay is expected given the total time necessary for human actions to be performed. In general, is expected that the delay of local processing would be lower than edge or cloud processing, due to the network delay that is added to the process. In opposite, the results illustrated in Fig.~\ref{fig:latency} shows that the action recognition time at the edge is approximately 50\% lower than processing locally. This occurs due to the lower processing power of the drone (even using the VPU). The VPU is not able to process the CNN as fast as the GPU at the edge. Besides, the network conditions are ideal, e.g., there is no background traffic, which contributes to the reduction of the delay using edge processing. More realistic scenarios involving heterogeneous background traffic will be evaluated in a future work.

\begin{figure}[htbp]
   \centering
   \includegraphics[width=0.45\textwidth]{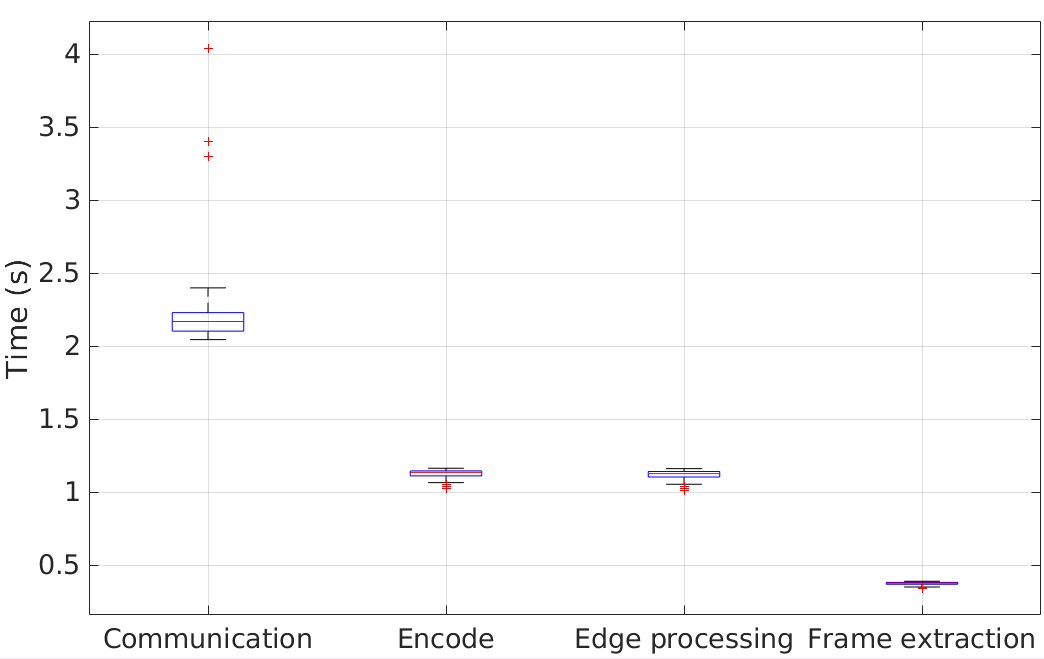}
   \caption{Edge processing time at different stages.}
   \label{fig:power_consumption}
\end{figure}

Besides the action recognition time at the edge, we evaluated the processing time of each stage involving the edge to determine where the delay is more critical. Analyzing Fig.~\ref{fig:power_consumption}, the network delay~\cite{baldi2000end} (i.e., the time to send and receive the data) is the most impacting, having an average about 2.2 seconds, approximately. The encoding delay, and the edge processing are about 1 second, and the frame extraction (from the camera video) delay is about 0.3 seconds. These information can be useful to know where to investigate methods to optimize the edge system. In other words, since the network delay is responsible for almost 50\% of the delay to recognize the action, developing methods to improve this communication will have more relevance for delay mitigation.

\begin{figure}[htbp]
   \centering
   \includegraphics[width=0.45\textwidth]{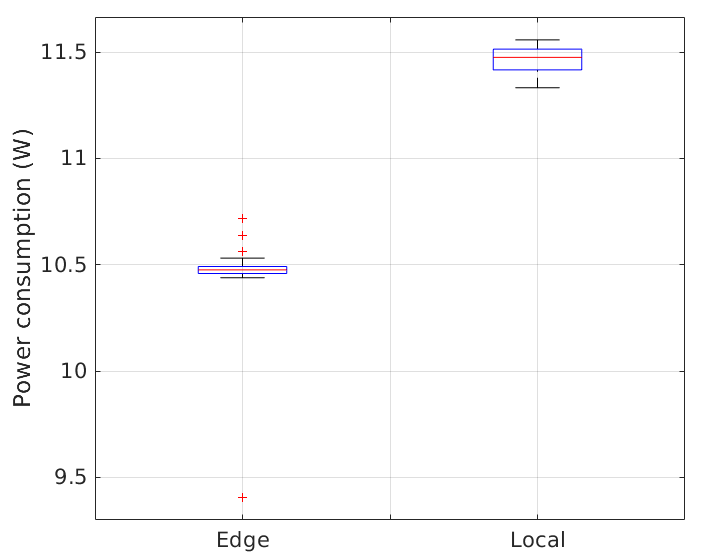}
   \caption{Power consumption of scenarios for each approach.}
   \label{fig:power_consumption2}
\end{figure}

Reduced power consumption is expected in the edge processing scenario due to the reduced number of operations performed in the UAV. The difference between both approaches is approximately 1~W, as shown in Fig.~\ref{fig:power_consumption2}. The local processing has an average power consumption about 11.43~W, and the edge processing is about 10.47~W.  

Equations~\ref{eq:results_01} and~\ref{eq:results_02} show the estimated battery lifetime for the local processing and edge processing, respectively.

\begin{equation}
  \frac{55.5 \: Wh}{11.43 \: W} = 4.86 \: h
  \label{eq:results_01}
\end{equation}

\begin{equation}
  \frac{55.5 \: Wh}{10.47 \:  W} = 5.30 \: h
  \label{eq:results_02}
\end{equation}

The battery lifetime using the edge is about 9\% greater than local processing. Therefore, edge processing can be more suitable for scenarios where network conditions are excellent and in situations where the battery lifetime and reaction time are critical requirements. 

It is important to notice that the given values of battery lifetime are not considering the rotors of the drone, therefore, in reality, they are smaller.

\section{CONCLUSIONS}
\label{sec:conclusions}

Real-time object detection relying on CNNs is computationally demanding and process it on a low-cost UAV platform is a challenging task. In this paper, we evaluated the impacts of processing CNN at UAV and in the edge, using an application running on top of OpenPose to detect the human gesture to control the drone. We verified through the results that edge processing is more efficient and faster than local processing when the network conditions are favorable. Moreover, we successfully developed a testbed that will drive our next work to evaluate more challenging scenarios, such as the split of CNN through the network. For future works, we intend to use more realistic scenarios, such as different levels of background traffic, increased number of UAVs (through emulation), and different kind of applications in the drone using different CNNs architectures. We also will compare the proposed algorithm to recognize if a person raised its arm with other solutions to prove its viability.

%\addtolength{\textheight}{-12cm}   % This command serves to balance the column lengths
                                  % on the last page of the document manually. It shortens
                                  % the textheight of the last page by a suitable amount.
                                  % This command does not take effect until the next page
                                  % so it should come on the page before the last. Make
                                  % sure that you do not shorten the textheight too much.

%%%%%%%%%%%%%%%%%%%%%%%%%%%%%%%%%%%%%%%%%%%%%%%%%%%%%%%%%%%%%%%%%%%%%%%%%%%%%%%%

%%%%%%%%%%%%%%%%%%%%%%%%%%%%%%%%%%%%%%%%%%%%%%%%%%%%%%%%%%%%%%%%%%%%%%%%%%%%%%%%

\section*{ACKNOWLEDGMENT}

This work was supported by the Innovation Center, Ericsson
Telecomunica\c c\~oes S.A, CNPq and Capes Foundation, Brazil.

\bibliographystyle{unsrt}
\bibliography{ref}

\end{document}